\documentclass[prd,preprint,superscriptaddress,amsmath,amssymb,nofootinbib]{revtex4}
\usepackage{graphicx,color,slashed}

\begin{document}

\title{Search for a light-charged Higgs  \\
in a two-Higgs-doublet type II seesaw model at the LHC}
\preprint{KIAS-P16070}
\author{Chuan-Hung Chen}
\email{physchen@mail.ncku.edu.tw}
\affiliation{Department of Physics, National Cheng-Kung University, Tainan 70101, Taiwan}

\author{Takaaki Nomura}
\email{nomura@kias.re.kr}
\affiliation{School of Physics, KIAS, Seoul 130-722, Korea}

\date{\today}

\begin{abstract}
A charged Higgs in the type II  two-Higgs-doublet model (THDM) has been bounded to be above a few hundred  GeV by the radiative $B$ decays. A Higgs triplet extension of the THDM  not only provides an origin of neutrino masses and a completely new doubly-charged Higgs decay pattern, but it also  achieves a light-charged Higgs with a mass of ${\cal O}(100)$ GeV  through the new scalar couplings in the scalar potential. It was found that these light-charged Higgs decays depend on its mass $m_{H^\pm}$, $\tan\beta$, and mixing effect $\sin\theta_\pm$: at $\tan\beta =1$, if $m_{H^\pm} >  m_W + m_Z$, $\bar b b W^\pm$, $W^\pm Z$, and $\tau \nu$ are the main decay modes;  however, if $m_{H^\pm} < m_W + m_Z$, the main decay modes are then $\bar b b W$ and $\tau \nu$, and  at $\tan\beta=30$, the $\tau \nu$ mode dominates the other decays. When $m_t > m_{H^\pm} + m_b$, we found that  the ATLAS and CMS recent upper bounds on the product of $BR(t\to H^+ b)BR(H^+\to \tau^+ \nu)$ can be directly applied and will give a strict constraint on the correlation of $m_{H^\pm}$ and $\sin\theta_\pm$.  If the upper bound of $BR(t\to H^+ b)BR(H^+\to \tau^+ \nu)$ is satisfied (escaped) for $m_t > (<) m_{H^\pm} + m_b$, it was found  that the significance of discovering the  charged Higgs through $H^\pm \to W^\pm Z$ is much lower than that through $H^\pm \to  \bar b b W^\pm$. With a luminosity of 100 fb$^{-1}$ at $\sqrt{s}=13$ TeV  and including the experimental bounds, the significance of the  $H^\pm \to \bar b b W^\pm$ signal can reach around $6.2 (2.4)\sigma$ for $m_{H^\pm} <(>)  m_W + m_Z$.

\end{abstract}

\maketitle

The two-Higgs-doublet model (THDM) is one of the minimal extensions of the standard model (SM) in supersymmetric (SUSY) and non-SUSY frameworks. The  scalar bosons in the model comprise two CP-even scalars $(h, \eta^0)$, one CP-odd pseudoscalar $(\chi^0)$, and two charged Higgs particles $(\eta^\pm)$. To avoid the flavor-changing neutral currents at the tree level, a discrete symmetry is usually imposed. Thus,   several types of non-SUSY THDMs have been classified in the literature according to the  couplings to the fermions in the Yukawa sector~\cite{Branco:2011iw,Ahn:2010zza};  As such, different types of THDMs may have different constraints on their masses and couplings.

With the discovery  of a new  scalar via ATLAS~\cite{:2012gk} and CMS~\cite{:2012gu},  the mass of the SM-like Higgs $h$ was determined to be  $m_h\approx 125$ GeV. Without fine-tuning the parameters in the scalar potential,  the $2\times 2$  mass-square matrix ($M^2$) elements of the CP-even scalars are expected to have the same order of magnitude. Due to the sizable off-diagonal entries of $M^2$, the mass-splitting between the two CP-even Higgs bosons can reach  a few hundred GeV. For instance, if the diagonal and off-diagonal elements of $M^2$ are $500^2$ GeV$^2$ and $300^2$ GeV$^2$, respectively, then the masses of the CP-even bosons would be  126 GeV and 583 GeV. 

Among the classifications of the model, only the type II THDM  has the same Yukawa couplings as the minimal supersymmetric standard model (MSSM). It is of importance that the charged Higgs  mass  of this model is bounded to be $m_{\eta^\pm}>480$ GeV at the $95\%$ confidence level (CL)~\cite{Misiak:2015xwa} by the precision measurement of $B\to X_s \gamma$~\cite{PDG}. Additionally, the constraint is insensitive to the parameter $\tan\beta = v_2/v_1$, where $v_{1(2)}$ is the vacuum expectation value (VEV) of the Higgs doublet that couples to the up(down)-type quarks. As a result, the charged Higgs in the type II THDM can not be a light particle unless the model is further extended. 

One of the unsolved puzzles in particle physics is the origin of neutrino masses. If we assume that the neutrino mass arises from spontaneous symmetry breaking (SSB), like that in the SM and THDM, the minimal extension of the THDM adds an $SU(2)$ Higgs triplet $\Delta$~\cite{ Magg:1980ut,Konetschny:1977bn}.  In addition to the new scalar bosons, such as doubly-charged Higgses $\delta^{\pm\pm}$ and neutral scalars $(\delta^0, \xi^0)$,  a pair of new charged scalar bosons $\delta^\pm$ exists in such models. Due to the strict constraint from the precision measurement of the electroweak $\rho$-parameter~\cite{PDG}, the VEV of $\Delta$ ($v_\Delta$) is limited to $v_\Delta < 3.4$ GeV. That is,   before electroweak symmetry breaking (EWSB), the triplet-scalar bosons are degenerated massive particles.  It can be easily seen that if only one Higgs doublet and one Higgs triplet are involved in a model, e.g. the so-called type-II seesaw model~\cite{ Magg:1980ut,Konetschny:1977bn}, the mass-splittings among the Higgs-triplet components are suppressed by $v_\Delta/m_{\Delta}$, where $m_\Delta$ is the typical mass of the Higgs triplet.  By assuming that  $\delta^{\pm \pm}$ are 100\% decayed into leptons, the experimental lower bound on the $m_\Delta$ is currently  around 400 GeV~\cite{Chatrchyan:2012ya, ATLAS:2012hi}.

Some interesting characteristics in the two-Higgs-doublet (THD) type II seesaw model have been reported~\cite{Chen:2014xva,Chen:2014qda}: (i) the mass-splittings among the triplet components can be of the $O(m_W)$; (ii) the doubly-charged Higgses $\delta^{\pm \pm}$ have a completely new dominant decay pattern, e.g., $\delta^{\pm\pm} \to H^\pm H^\pm$, where $H^\pm$ are the lightest singly charged Higgses; and, (iii) the $H^\pm$ can be as light as ${\cal O}(100)$ GeV. These characteristics are ascribed to the new interactions in the scalar potential, which are summarized   as~\cite{Chen:2014xva,Chen:2014qda}:
 \begin{align}
V(\Phi_1,\Phi_2,\Delta) & \supset   \mu_1 \Phi^T_1 i\tau_2 \Delta^{\dagger}  \Phi_1 + \mu_2 \Phi^T_2 i \tau_2 \Delta^\dagger \Phi_2 + \mu_3 \Phi^T_1 i\tau_2 \Delta^\dagger \Phi_2 + h.c. \,, \label{eq:vPhiDelta}
 \end{align}
where $\Phi_{1,2}$ are the Higgs doublets; $\tau_2$ is the second Pauli matrix, and $\mu_{1,2,3}$ are the mass dimension-one parameters and can be on the order of an electroweak scale. The VEV of $\Delta$  can then be simplified as:
 \begin{equation}
  v_\Delta \sim \frac{1}{\sqrt{2}m^2_\Delta} \left( \mu_1 v^2_1 + \mu_2 v^2_2 + \mu_3 v_1 v_2 \right)\,, \label{eq:v_d}
  \end{equation}
 the detailed expression of which  can be found in~\cite{Chen:2014xva}. As a result, the small $v_\Delta$ can be achieved by taking proper values of $\mu_{1,2,3}$. According to the results of $\mu_{1,2,3}\sim {\cal O}(m_W)$, the off-diagonal singly-charged Higgs mass-square matrix can be compatible with  $m^2_{\eta^\pm}$ and $m^2_\Delta$; and  following the earlier discussion on the case of  the two CP-even bosons,  one indeed can obtain the light-charged Higgs with a mass of 100 GeV, even  $m_\Delta \sim m_\eta^\pm \sim 500$ GeV. The only consequence is to introduce a mixing effect between $\eta^\pm$ and $\delta^\pm$.  Since  the doubly-charged Higgs  of  the Higgs triplet does not mix with other particles, its mass is $m_{\delta^{\pm \pm}}  \approx  m_\Delta$. A detailed analysis for $\delta^{\pm\pm}$ production at the LHC in this model can be found in~\cite{Chen:2014qda}. 

Direct searches for a light-charged Higgs were performed at the LHC with  $\sqrt{s}=$7 TeV~\cite{Aad:2012tj,Aad:2012rjx,Aad:2013hla,Chatrchyan:2012vca} and 8 TeV~\cite{Aad:2014kga, Khachatryan:2015uua,Khachatryan:2015qxa}.  Although  significant events above the background have not  yet been found, the search for a light-charged Higgs still continues  and  remains an interesting issue at the LHC~\cite{Akeroyd:2016ssd,Akeroyd:2016ymd,Arhrib:2016wpw}. Based on the characteristics  of  the THD type II seesaw model, in this work, we study the possible signatures of the  light-charged Higgs at the LHC.  Since the triplet-charged Higgses $\delta^\pm$ can couple to $W^\mp Z$ at the tree level,  the decays of $H^\pm \to W^\pm Z$ are sizable when $m_{H^\pm} > m_W + m_Z$; therefore, we separated the light-charged Higgs mass into two ranges , namely (I) $m_W+ m_Z \leq m_{H^\pm} < m_t + m_b$ and (II) $m_{H^\pm} < m_W + m_Z$, where the light-charged Higgs predominantly decays to $\bar b b W$, $WZ$,  $\tau \nu$, and $cs$ in the former range and decays to $\bar b b W$, $\tau \nu$, and $cs$ in the latter range. 
  It is worth mentioning  that if the pseudoscalar $A^0$ of the THD is lighter than the charged Higgs,  the $H^\pm \to W^\pm A^0$ decay may become the dominant decay channel. 
Since we do not have any information about the mass of the pseudoscalar boson ($A^0$), we omit  the discussions of the  $H^\pm \to W^\pm A^0$ decays by assuming $m_{A^0} > m_{H^\pm}$ in the study range of $m_{H^\pm}$. This assumption is supported by the recent CMS measurement,  in which the mass $m_{A^0}$ in the range $20-270$ GeV at small $\tan\beta$ was excluded~\cite{Khachatryan:2016are}. 
Although the $H^\pm \to W^\pm h$  decays could be interesting processes  to investigate   the charged Higgs, the vertices are suppressed by $\cos(\beta-\alpha)$, which has been strictly limited by the current Higgs data~\cite{Benbrik:2015evd}. While it is not necessary, for simplicity, we adopt $\cos(\beta-\alpha) \approx 0$  in this analysis.

In order to study the light-charged Higgs production and decays, we briefly introduce the relevant mixing and couplings in the following. As discussed earlier, the singly-charged Higgses, namely  $\eta^\pm$  and  $\delta^\pm$, can mix together due to the interactions in Eq.~(\ref{eq:vPhiDelta}). We parameterize the physically-charged Higgs states as: 
 \begin{equation}
  \left( \begin{array}{c}
    H^{'\pm}_1\\ 
    H^\pm \\ 
  \end{array}\right) =   \left(\begin{array}{cc}
    \cos\theta_\pm & \sin\theta_\pm \\ 
    -\sin\theta_\pm & \cos\theta_\pm\\ 
  \end{array}\right) \left( \begin{array}{c}
    \eta^\pm \\ 
    \delta^{\pm} \\ 
  \end{array}\right)\,,\label{eq:ma}
 \end{equation}
where $H^\pm (H'^\pm)$  are identified as the lighter (heavier) charged Higgses. Accordingly, the Yukawa couplings of $H^\pm$ to the SM quarks are given by:
\begin{align}
{\cal L}^{H^\pm}_Y &= -\sqrt{2}\sin\theta_\pm  \bar u  \left[  \frac{ t_\beta}{v }{\bf V}{\bf m_d} P_R  + \frac{ 1}{v t_\beta } {\bf m_u}  {\bf V}  P_L \right] d H^+  + h.c.\,, \label{eq:qyu}
%
\end{align}
in which we have suppressed the flavor indices, and where  ${\bf V}$ denotes the Cabibbo-Kobayashi-Maskawa (CKM) matrix, and $t_\beta =\tan\beta$.  One can easily obtain the lepton Yukawa couplings if the quark masses and CKM matrix are respectively replaced by the lepton masses and Pontecorvo-Maki-Nakagawa-Sakata (PMNS) matrix.  It is clear that the radiative $B$ decaying via the charged Higgs can be  suppressed by $\sin^2\theta_\pm$ in the decay amplitude; in this manner,  the constraint on the light-charged Higgs  is  relaxed.  We then write the Feynman rules for the gauge interactions as: 
\begin{align}
     Z  \chi^+ \chi^- & : \frac{ g c_{\chi}}{2c_W} (p^\mu_{+} - p^\mu_{-})\,, \nonumber \\
    \quad A \chi^+ \chi^- & :  e  (p^\mu_{+} - p^\mu_{-})\,, \nonumber \\
    WZ\delta^\pm & :-\frac{ g^2 v_\Delta}{\sqrt{2} c_W}  g^{\mu\nu}\,,
\end{align}
where  we have used the basis of $\eta^\pm$ and $\delta^\pm$ instead of $H^\pm$ and $H'^\pm$, $\chi^\pm$  represent  $\eta^\pm$ and $\delta^\pm$,  and $p^\mu_\pm$ denote the momenta of the charged particles,  $c_W(s_W)=\cos\theta_W(\sin\theta_W)$,  $c_{\eta}= 1-2s^2_W$, and $c_\delta = -s^2_W$. It can be seen that the $W^- Z H^+$ coupling is suppressed by $v_\Delta/v$; however, when  $t_\beta \sim {\cal O}(1)$, its magnitude is compatible with the $m_\tau t_\beta /v$ coupling of  $\tau \nu H^+$. Thus, in the (I) mass range, the relative branching ratios (BRs) for $H^\pm \to W^\pm Z$ and $H^\pm \to \tau \nu$ are sensitive to the $\sin\theta_\pm$. 

After introducing  the relevant couplings,  we then study  the partial decay rates of the charged Higgs for the kinematically-allowed channels. It was found that in the  (I) mass range, the partial-decay rate indeed is dominated by the three-body decay $H^+ \to \bar b t^* \to \bar b  bW^+$, the expression of which  is written  as:
 \begin{align}
\Gamma(H^\pm \to \bar b b W^\pm) & \approx \frac{N_c g^2 s^2_\pm }{2 m_{H^\pm} } \int^{(m_{H^\pm} -m_b)^2}_{(m_W+m_b)^2} \left| \frac{1}{q^2 -m^2_t -i \Gamma_t m_t}\right|^2 \nonumber \\
& \times 
 \left[ \left( \frac{m^2_b t^2_\beta}{v^2}+ \frac{m^2_t}{q^2} \frac{m^2_t }{v^2 t^2_\beta} \right) p_H\cdot p_{b1} 
 + 2 \frac{m^2_b m^2_t}{v^2} \right] \nonumber \\
 & \times \left[ 2p_t\cdot p_{b2} + \frac{1}{m^2_W} \left( (q^2 -m^2_b)^2 -m^4_W\right) \right] d_3(PS)\,,
\end{align}
where $p_H\cdot p_{b1}$ and $p_t\cdot p_{b2}$ are the inner products of the particle momenta,  and $d_3(PS)$ denotes the phase space factor of the three-body decay. In sum, they are given by:
\begin{align}
p_H\cdot p_{b1} & = \frac{1}{2} \left( m^2_H -q^2 -m^2_b\right)\,, \quad p_t\cdot p_{b2} = \frac{1}{2} \left(q^2 -m^2_W +m^2_b \right)\,, \nonumber\\
d_3(PS) & = \frac{dq^2 }{(2\pi)^5} \left( \frac{\pi}{m_{H^\pm}} \sqrt{E^2_{b1} - m^2_b}\right) \left(\frac{\pi}{\sqrt{q^2}} \sqrt{E^2_W -m^2_W}\right) \,, \nonumber \\
E_{b1} & = \frac{m^2_{H^\pm} - q^2  +m^2_b}{2 m_{H^\pm}}\,, \quad E_W = \frac{q^2 +m^2_W -m^2_b}{2\sqrt{q^2}}\,.
\end{align}
Meanwhile,  the dominant  two-body decays are formulated as:
\begin{align}
\Gamma(H^\pm \to \tau  \nu) & \approx \frac{m_{H^\pm}}{8\pi}\left(  s_{\pm} \frac{m_\tau t_\beta}{v}\right)^2 \left( 1 - \frac{m^2_\tau}{m^2_{H^\pm}}\right)^2 \,, \nonumber \\
\Gamma(H^\pm \to c s) & \approx \frac{N_c s^2_{\pm} m_{H^\pm} }{8\pi}\left( \frac{m^2_s t^2_\beta}{v^2} + \frac{m^2_c }{v^2 t^2_\beta} \right) \left( 1 - \frac{m^2_c}{m^2_{H^\pm}}\right)^2\,, \nonumber \\
\Gamma(H^\pm \to W^\pm Z) & \approx \frac{g^4 c^2_\pm v^2_\Delta}{32 \pi c^2_Wm_{H^\pm}}  \sqrt{\lambda(r_Z, r_W)} \left[2+ \frac{(m^2_{H^\pm} -m^2_Z -m^2_W )^2}{4m^2_Z m^2_W} \right]\,,
%
\end{align}
in which  the light fermion masses were dropped, and  $V_{cs}\approx 1$ is used;  and where $s_\pm (c_\pm) =\sin\theta_\pm (\cos\theta_\pm)$; $r_Z=m^2_Z/m^2_{H^\pm}$, $r_W=m^2_W/m^2_{H^\pm}$, and $\lambda(a,b)=(1+a-b)^2 -4 a$.  
In addition to  the $\tau \nu$ and $cs$ modes, in the (II) mass range  one of the vector bosons in the vector-boson pair channels becomes off-shell. Although the contributions of the off-shell channels are small, their partial decay rates are still expressed as follows:
\begin{align}
\Gamma(H^\pm \to  Z W^{\pm*}) &\approx \frac{3 g^6 v^2_{\Delta} c^2_{\pm}}{ 2^{8} \pi^3 c^2_{W}} \int^{\Delta M^2_Z}_0 dq^2 \frac{q^2 \sqrt{\lambda(r_Z, r_{q})}}{|q^2 -m^2_W|^2}\left[1+ \frac{E^2(m^2_Z,q^2)}{2m^2_Z}\right]\,, \nonumber \\
\Gamma(H^\pm \to  W^\pm Z^*) &\approx \frac{ g^6 v^2_{\Delta} c^2_{\pm} \xi_{VA} }{ 9\cdot 2^{8} \pi^3 c^4_{W}} \int^{\Delta M^2_W}_0 dq^2 \frac{q^2 \sqrt{\lambda(r_W, r_{q})}}{|q^2 -m^2_Z |^2} \left[ 1 + \frac{E^2(m^2_W,q^2)}{2 m^2_W}\right]
\end{align}
where $\Delta M^2_V = \left( m_{H^\pm} -m_V \right)^2$,  $\xi_{VA}=63/2+20s^2_W (4 s^2_W -3) $, and $E(m^2, q^2)=(m^2_{H^\pm} - m^2 -q^2)/(2\sqrt{q^2})$. 

\begin{figure}[hptb] 
\includegraphics[width=80mm]{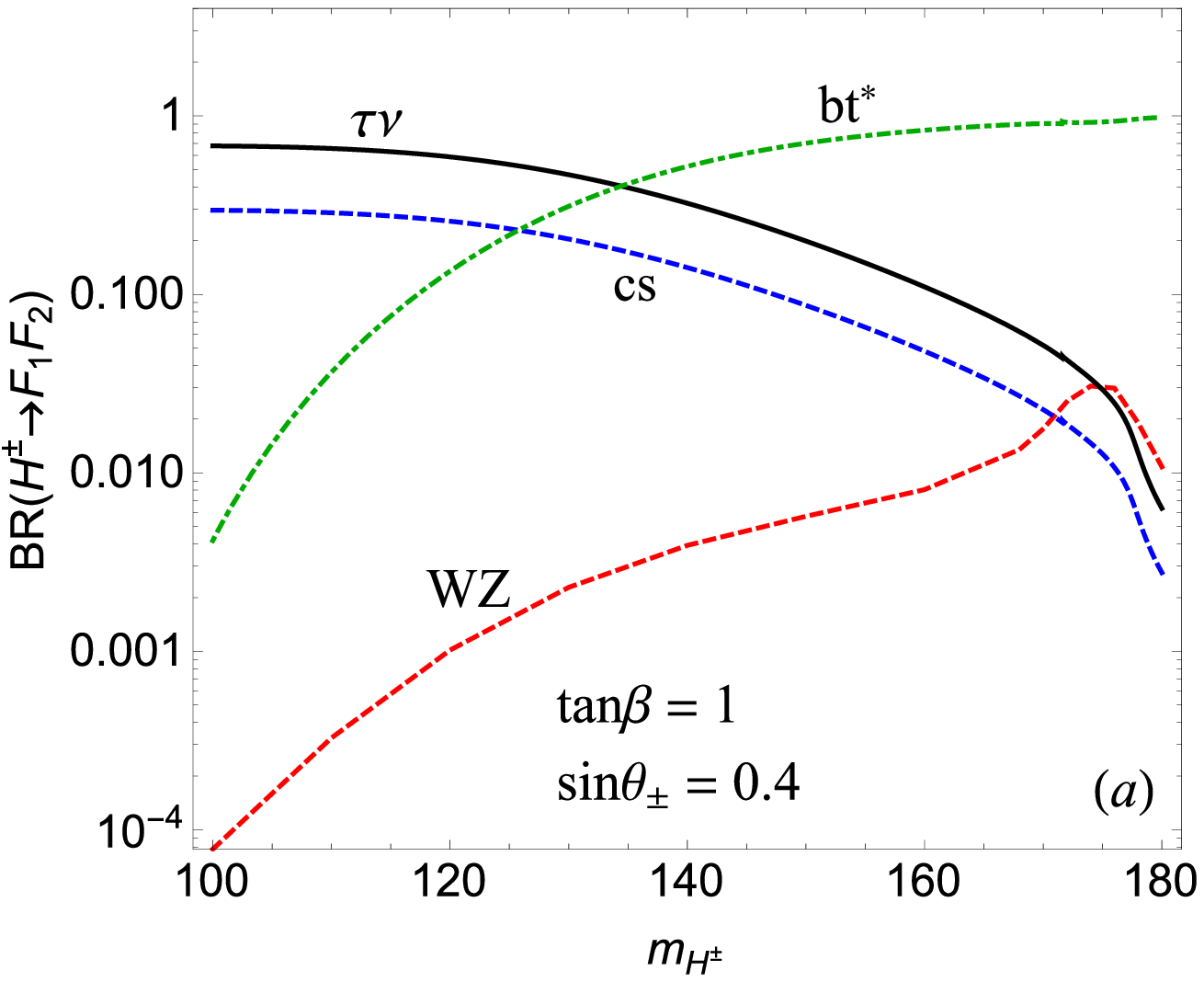}
\includegraphics[width=80mm]{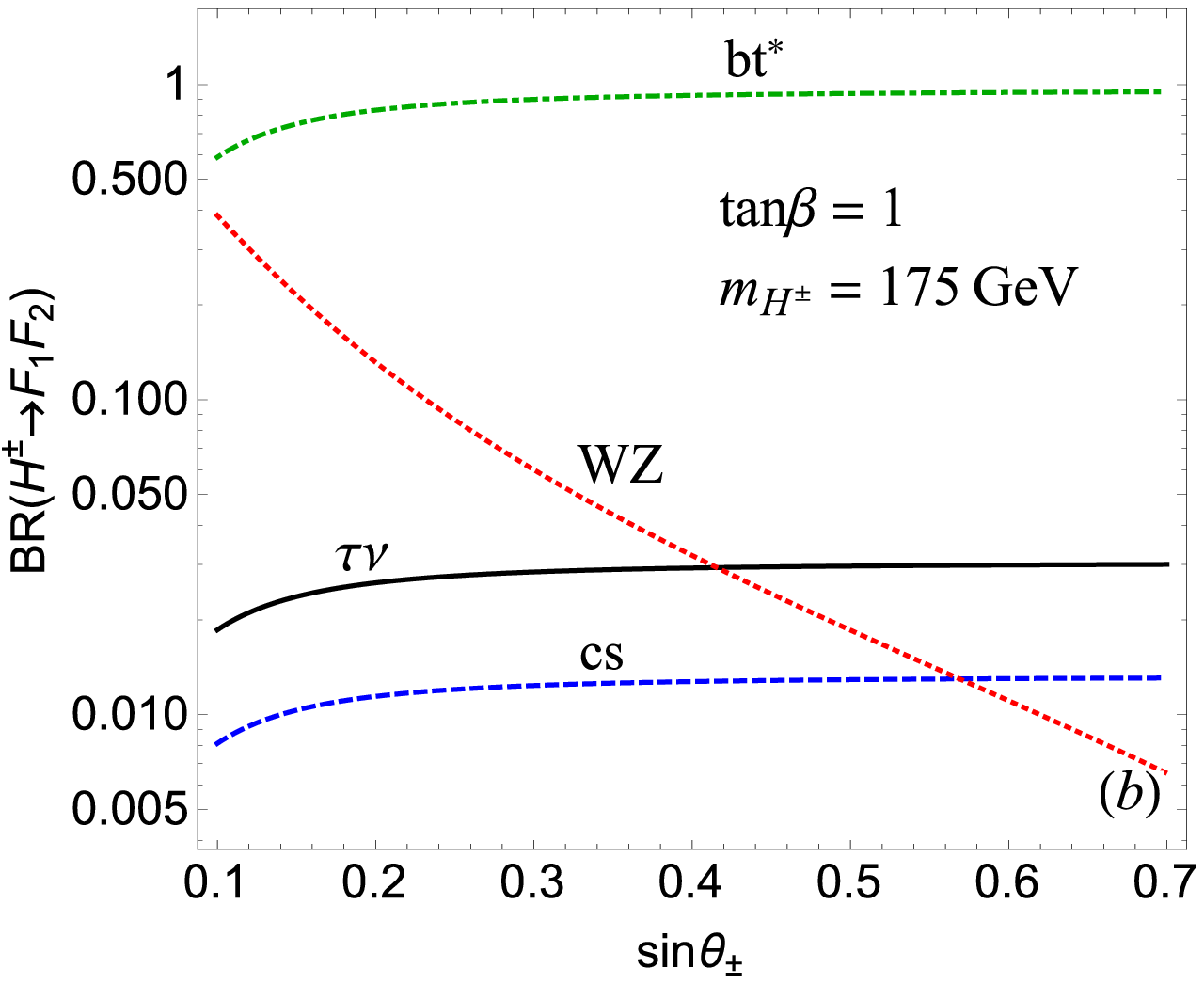} 
\caption{ BRs for the light-charged Higgs decays as a function of (a) $m_{H^\pm}$ and (b) $\sin\theta_\pm$, where  $\tan\beta=1$ for both plots,  $bt^*$ stands for  $\bar b b W$, $F_{1,2}$ denote the possible final states, and $\sin\theta_\pm =0.4$ for plot (a), while $m_{H^\pm}=175$ GeV for plot (b).} \label{fig:BRCH}
\end{figure}

According to  the obtained formulas for the $H^\pm$ decays, we show the BR for each $H^\pm$ decay mode as a function of $m_{H^\pm}$ in Fig.~\ref{fig:BRCH}(a) and as a function of $s_\pm$ in Fig.~\ref{fig:BRCH}(b), where $s_\pm=0.4$ and  $m_{H^\pm}=175$ GeV are used in plots (a) and (b), respectively,  $t_\beta=1$ is fixed in both plots,  and  $F_{1,2}$ in the y-axis denote the possible final states. Other taken values of parameters  are shown in Table~\ref{tab:inputs}. Without further elaboration,  $v_\Delta=3$ GeV is indicated in this work.
 Since the $W/Z$ gauge boson can be off-shell and on-shell in the $H^\pm$ decays when $m_{H^\pm}$ is taken as a variable,  to include the effects of the $W/Z$-gauge boson width,  the $BR(H^\pm \to  W^{\pm(*)} Z^{(*)})$ is calculated by summing  over all possible $W/Z$ decays, i.e. $BR(H^\pm \to W^{\pm(*)} Z^{(*)})=\sum_{f_1,f_2,f_3,f_4} BR(H^\pm \to f_1 f_2 f_3 f_4)$, where $f_is$  denotes all possible final states. Here, we employ CalcHEP~\cite{Belyaev:2012qa} with a narrow-width approximation to estimate the numerical values for the $BR(H^\pm \to W^{\pm(*)} Z^{(*)})$. For simplicity, we hereafter use $WZ$ instead of $W^{\pm(*)} Z^{(*)}$. 

   \begin{table}[hpbt]
 \begin{ruledtabular}
\begin{tabular}{ccccccc} 

$m_t (pole)$ & $m_b(m_b)$ & $m_W$ & $m_Z$ & $\Gamma_t$ & $\Gamma_W$ & $\Gamma_Z$   \\ \hline
 173  & 4.2 & 80.39 & 91.19 & 1.41 & 2.11 & 2.52 \\

\end{tabular}
\caption{ Inputs of particle masses and widths in units of GeV. }
\label{tab:inputs}
\end{ruledtabular}
\end{table}

\begin{figure}[hptb] 
\includegraphics[width=80mm]{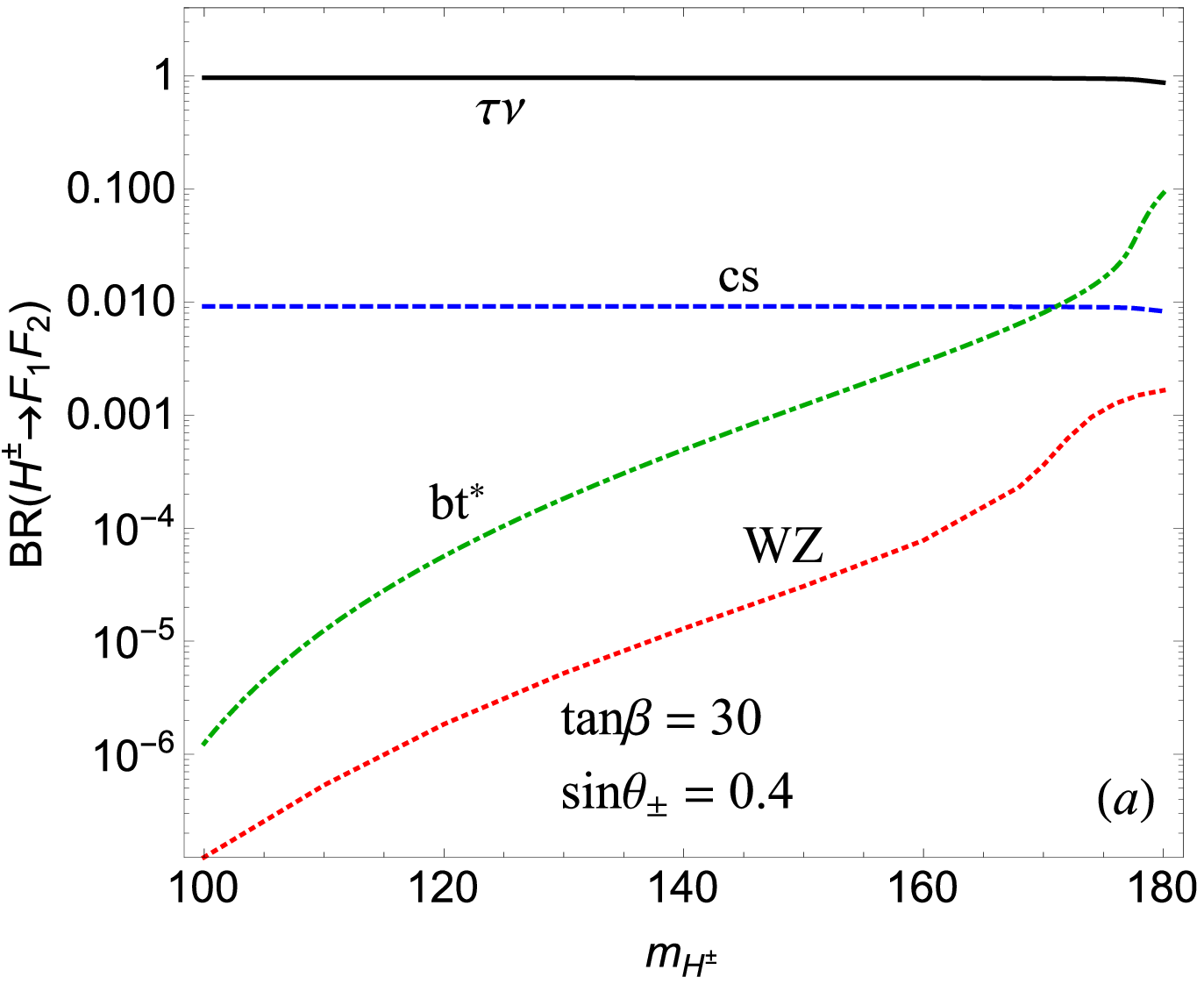}
\includegraphics[width=80mm]{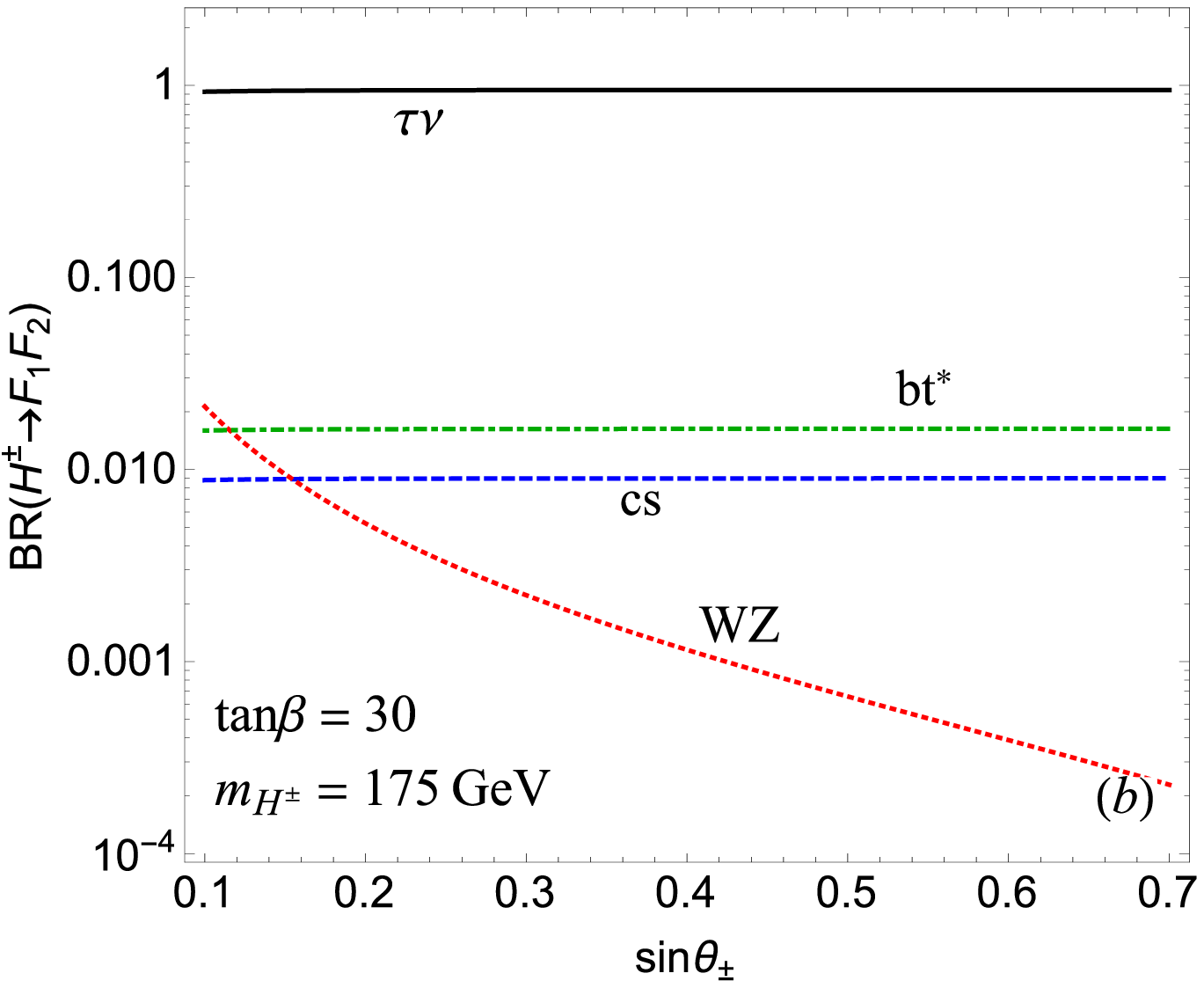} 
\caption{ This legend is the same as  that in Fig.~\ref{fig:BRCH}, except here  $\tan\beta=30$.} \label{fig:BRCH_tb30}
\end{figure}

From plot (a), it can be clearly seen that when $m_{H^\pm} > 160$ GeV, the $BR(H^\pm \to \bar b b W^\pm)$ is one order of magnitude larger than the other decay modes. The $BR(H^\pm \to \tau \nu)$ is about two-fold larger than the $BR(H^\pm \to cs)$ in both the  (I) and (II) mass ranges.  When $m_{H^\pm} > m_W + m_Z$, the BRs of the $\tau \nu$ and $W Z$ decay channels are compatible. From plot (b), it can be seen that  the $BR(H^\pm \to W Z)$  is enhanced when $s_\pm$ is decreased.  On the other hand, since $s_\pm$ dictates the single-charged Higgs  production cross-section, in order to  produce $H^\pm$ bosons with sizable cross-sections, the values of $s_\pm$ must not be too small. In order to clarify the situation with large values of $t_\beta$, we present the results with $t_\beta=30$ in Fig.~\ref{fig:BRCH_tb30}. From the results, it can  clearly be seen that the $\tau \nu$ mode overwhelmingly dominates with large $t_\beta$. 

\begin{figure}[hptb] 
\includegraphics[width=80mm]{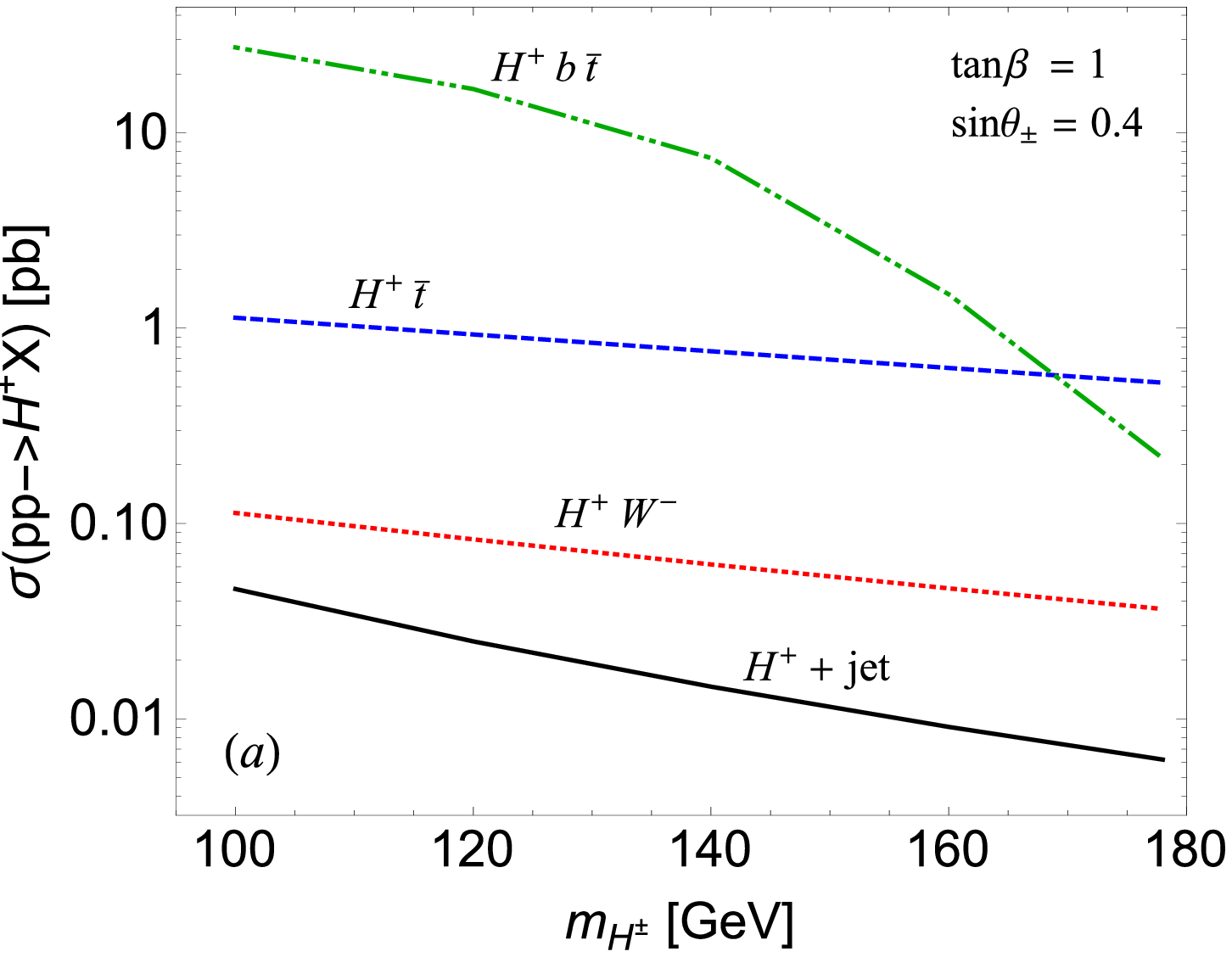}
\includegraphics[width=80mm]{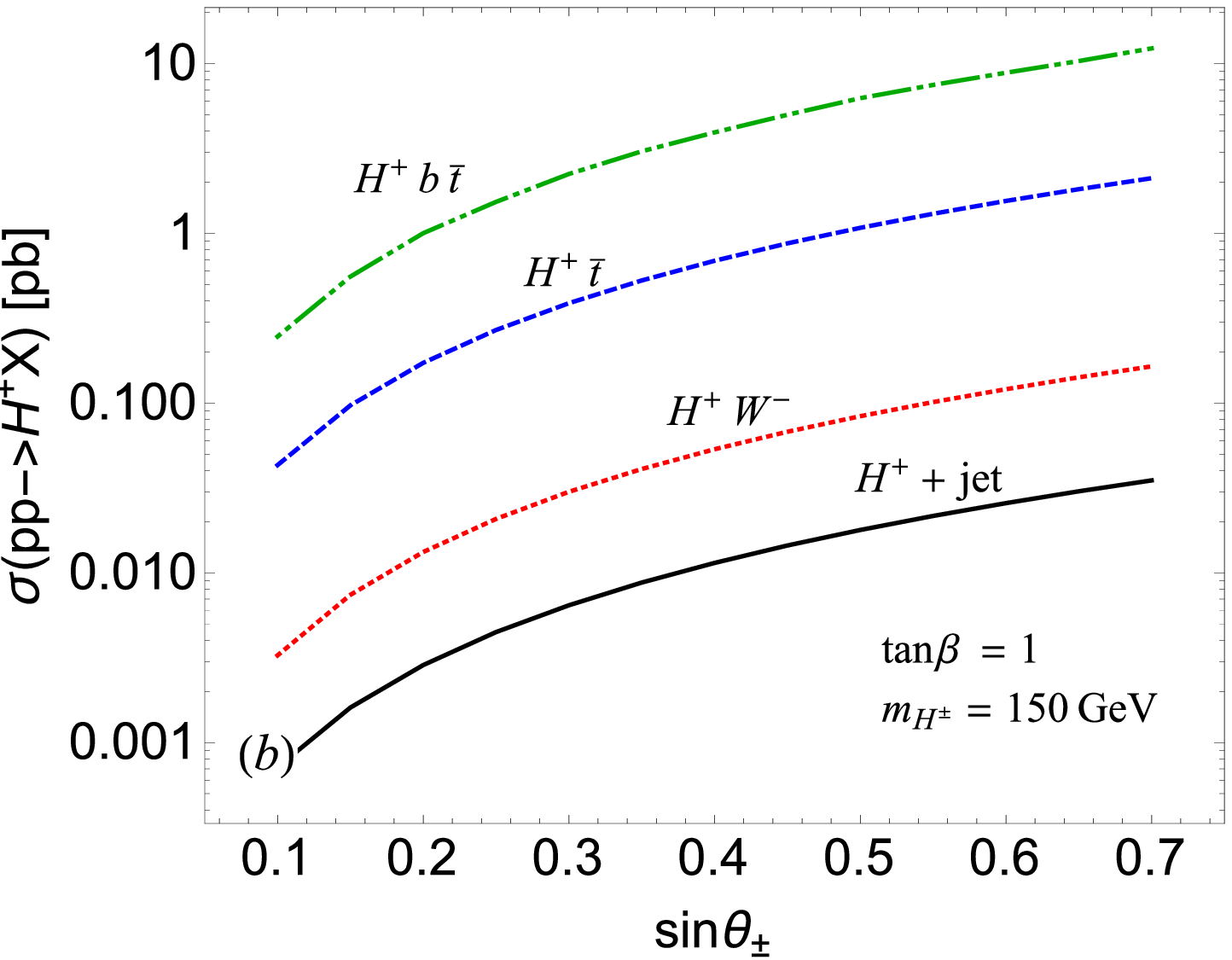} 
\caption{ Single-charged Higgs production cross-section at parton level via various channels as a function of (a) $m_{H^\pm}$ and (b) $\sin\theta_\pm$, where $\sqrt{s}=13$ TeV is used, we fixed $\tan\beta=1$ for both plots, and $\sin\theta_\pm =0.4$ for plot (a), while $m_{H^\pm}=150$ GeV for plot (b).} \label{fig:X_CH}
\end{figure}

To numerically calculate the single-charged Higgs production cross-section, we use CalcHEP~\cite{Belyaev:2012qa} associated with  the {\tt CTEQ6L} parton distribution functions (PDFs)~\cite{Nadolsky:2008zw}. The numerical analysis is conducted  at the center of the mass energy of $\sqrt{s}=13$ TeV. It was found that there are four channels of interest that could produce the single-charged Higgs in the $pp$ collisions. They are: $H^+ b \bar t$, $H^+\bar t$, $H^+ W^-$, and $H^+ +$ jet, where the CP-conjugated processes are indicated, and the jet includes the gluon, light quarks, and $b$ jets;  further, their respective producing processes respectively are: $gg \to  t \bar t \to H^+ b \bar t$, $g \bar b \to H^+ \bar t$, $q \bar q \to Z \to H^+ W^-$, and $g q \to H^+ q'$ ($q,q' = u, d, s, c$). The main free parameters in this study are $t_\beta$, $s_\pm$, and $m_{H^\pm}$. To show the correlations between the cross-sections and $m_{H^\pm}$,  the production cross-sections of these channels are ploted as a function of $m_{H^\pm}$ in Fig.~\ref{fig:X_CH}(a), in which  $t_\beta=1$ and $s_\pm =0.4$. According to the results, we see that the cross-sections of the $H^+ b\bar t $ and $H^+ \bar t $ channels are much larger than those of the $H^+ W^- $ and $H^+ +$ jet channels. Similarly, the production cross-sections  are shown as a function of $s_\pm$ in Fig.~\ref{fig:X_CH}(b), for which $t_\beta =1$ and $m_{H^\pm}=150$ GeV. It can be seen that when the values of $s_\pm$ are around 0.2, both $\sigma(pp\to H^+ b \bar t)$ and $\sigma(pp\to H^+\bar t)$ can still exceed 100 fb. From the results shown in Fig.~\ref{fig:BRCH}(b), where  the $BR(H^\pm \to W Z)$ is larger than the $BR(H^\pm \to \tau \nu)$ in the (I)  mass region and small $s_\pm$, it is of interest to explore the charged Higgs through the $WZ$ channel in such regions. 

\begin{figure}[hptb] 
\includegraphics[width=80mm]{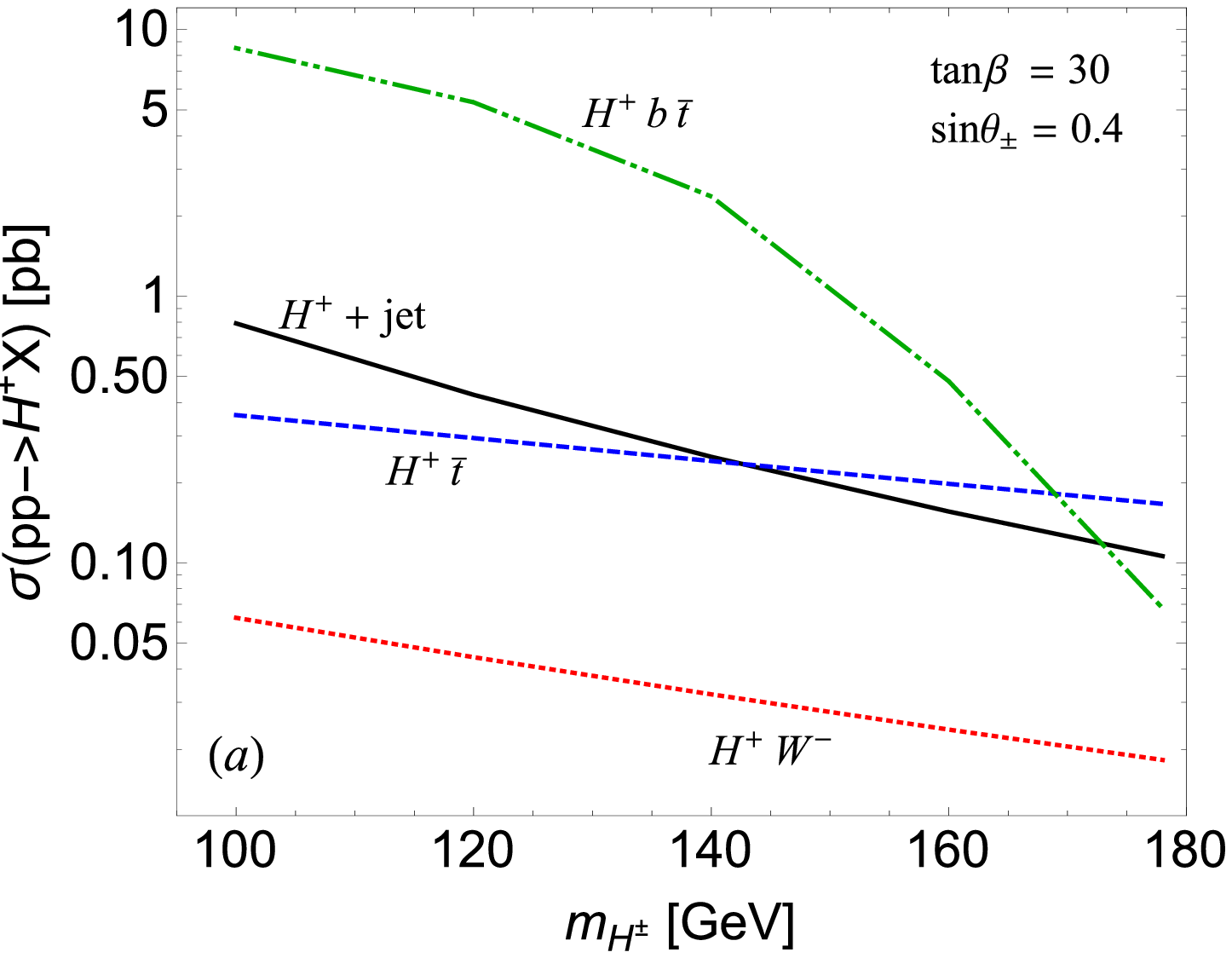}
\includegraphics[width=80mm]{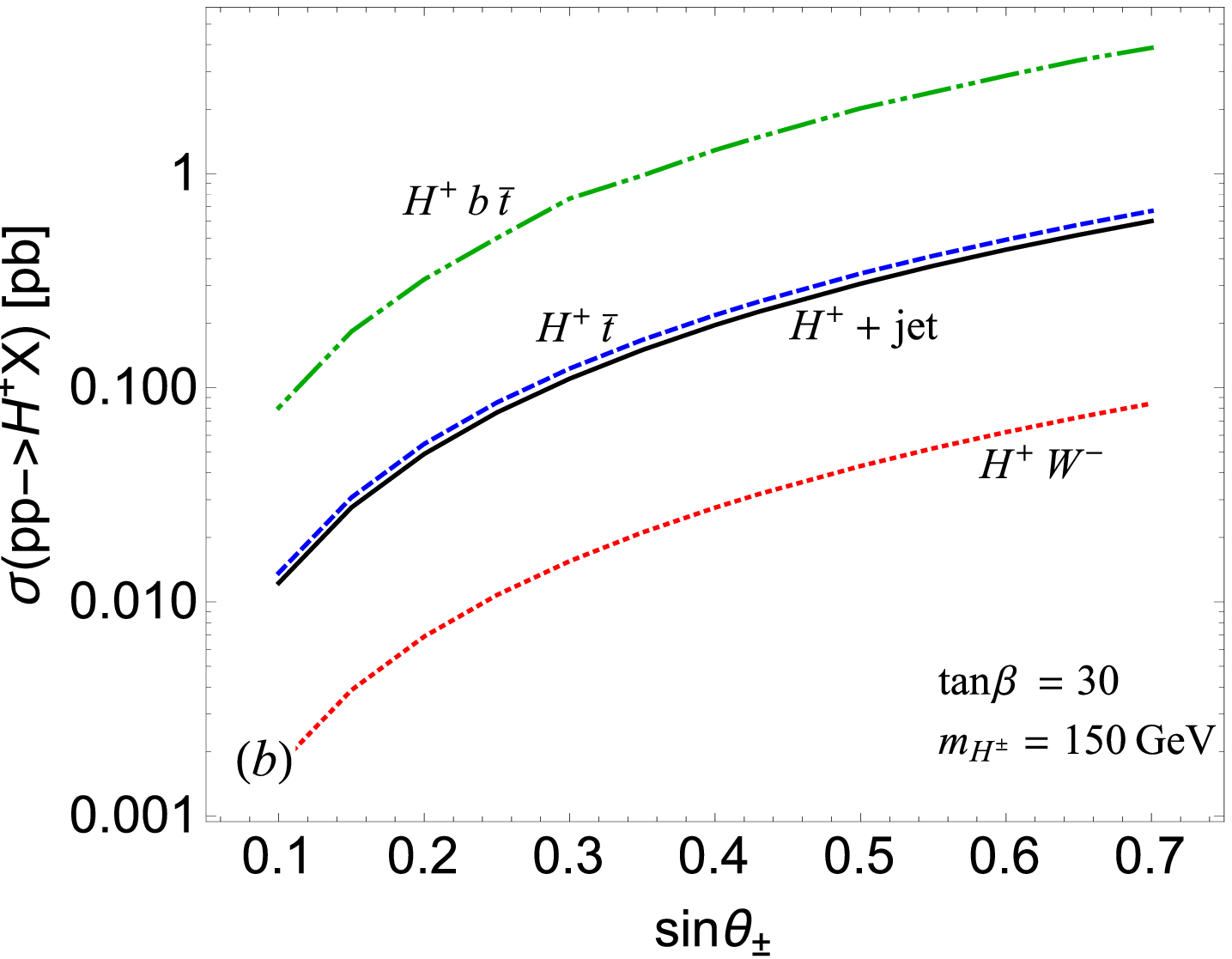} 
\caption{ This legend is the same as that in Fig.~\ref{fig:X_CH}, except here $\tan\beta=30$.} \label{fig:X_CH_tb30}
\end{figure}

According to the charged Higgs-Yukawa couplings shown in Eq.~(\ref{eq:qyu}), when $t_\beta$ increases, the contributions from up(down)-type quarks decrease (increase). Since the $t(c)$-quark is much heavier than the $b(s)$-quark, it is expected that the single $H^\pm$ production cross-sections would lower as the values of $t_\beta$ initially augment. However, when the values of $t_\beta $ increase up to $m_t/m_b$, due to $m_b t_\beta$ being near $m_t$ and $m_s t_\beta$ being larger than $m_c$, the cross-sections are enhanced close to the case  of $t_\beta=1$. To illustrate the situation with a large $t_\beta$, we show the $H^+$ production cross-sections as a function of $m_{H^\pm}$ and $s_\pm$, with $t_\beta=30$ in Figs.~\ref{fig:X_CH_tb30}(a) and (b), respectively. The plots show that only the $H^+ +$ jet channel has a significant enhancement, while  the others are slightly smaller than when $t_\beta=1$. The significant enhancement in the $H^+ +$ jet channel is due to the result of  $m_s t_\beta > 2 m_c$. 

 In order to assess the discovery potential of the light-charged Higgs, we employ the event generator  {\tt MadGraph5\_aMC@NLO/MadEvent5}~\cite{Alwall:2014hca} to generate  the simulation events,  where  the relevant Feynman rules and parameters of the model are provided by FeynRules 2.0 \cite{Alloul:2013bka}, and  the {\tt NNPDF23LO1} PDFs~\cite{Deans:2013mha} are used.  The generated events are passed through  {\tt PYTHIA\,6}~\cite{Ref:Pythia} to include the effects of hadronization and initial/final state radiation, and deal with the SM particle decays, e.g.,  the $W$-boson and top-quark decays. 
Furthermore,  the events are also run through  the  {\tt PGS\,4} for detector simulation~\cite{Ref:PGS}.

As aforementioned, the singly-charged Higgs $H^\pm$ is predominantly produced by the processes $pp \to H^\pm t b$, $pp \to H^\pm t$, and $pp \to H^\pm+$ jet, while the production cross-sections and decays depend on the parameters $m_{H^\pm}$, $s_\pm$, and $t_\beta$. Therefore, to combine the charged Higgs (I, II) mass ranges  with the  $t_\beta=(1, 30)$ schemes, we classify the schemes as $S_{1A,1B}$ and $S_{2A,2B}$, where the number  in the subscript denotes the mass range, and $A(B)$ represents  $t_\beta=1(30)$. 
{We note that it encounters a double counting if we simply add up the events  from  the $pp \to H^\pm t$ and $pp \to H^\pm t b$ processes, where the  events from the latter process may include those from the former process when the $b$-jet in the final state is collinear with the proton beam~\cite{Alwall:2004xw}.
 In order to remove the double counting effects, we  apply MLM matching scheme~\cite{Hoche:2006ph,Mangano:2006rw} which is implemented in {\tt MadGraph5\_aMC@NLO}. 
 It is found  that with $m_{H^\pm} =175$ GeV, the matched cross-section for  the $pp\to H^\pm t$ and $pp\to H^\pm t b$ channels is around $90 \%$ of that  obtained by a simple sum of both channels. In addition, we also find that the double counting effects  become weaker when the mass of charged Higgs is smaller. This behavior is consistent with that shown in~\cite{Alwall:2004xw}.  Although we concentrate on the case of $m_{H^\pm} < m_t + m_b$, in order to avoid the double counting issue, we use the events which are generated by MadGraph with MLM matching procedure in the following analysis.}

\begin{figure}[hptb] 
\includegraphics[width=80mm]{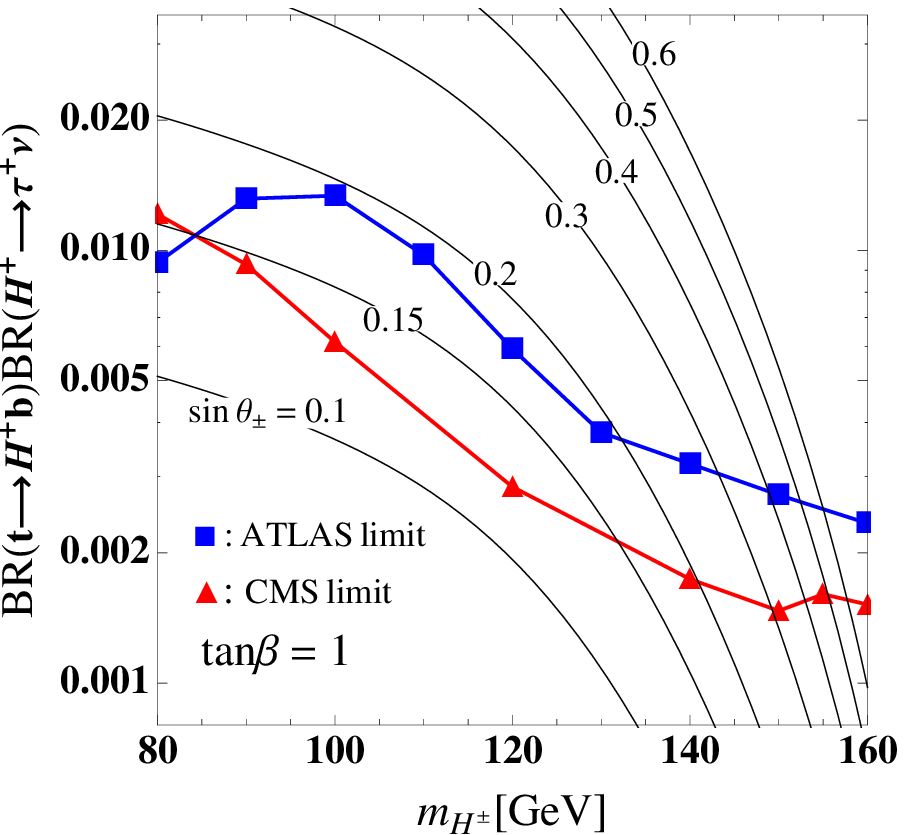}
\includegraphics[width=80mm]{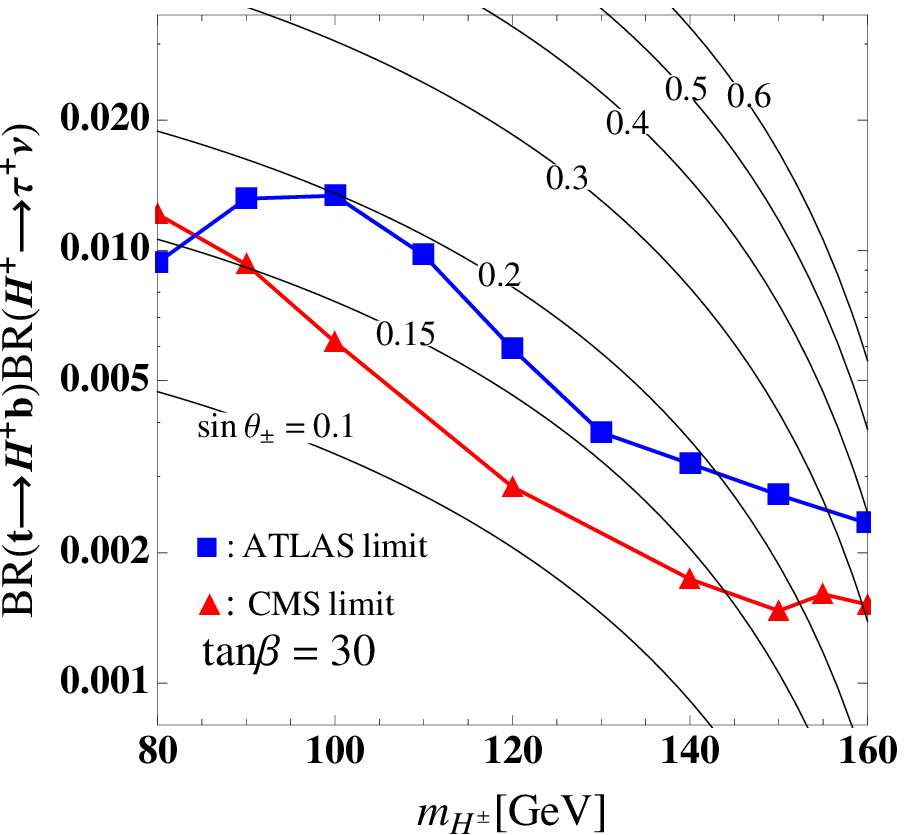} 
\caption{ Product of branching ratios $BR(t \to H^+ b)BR(H^+ \to \tau^+ \nu)$ as a function of $m_{H^\pm}$, where  the upper limits from the ATLAS~\cite{Aad:2014kga} and  CMS~\cite{Khachatryan:2015qxa} measurements are included;  the curves represent the different values of $s_\pm$, and the left (right) panel is for $\tan\beta=1 (30)$. } \label{fig:tdecay}
\end{figure}

From Fig.~\ref{fig:BRCH_tb30}, it can be seen that  the decay $H^\pm \to \tau \nu$ dominates the other decay modes at large values of $t_\beta$; that is, under such circumstances, $\tau \nu$ is the only channel that can be used to look for  $H^\pm$. Recently, ATLAS~\cite{Aad:2014kga} and CMS~\cite{Khachatryan:2015qxa} reported the upper limits on the product of $BR(t \to H^+ b)BR(H^+ \to \tau^+ \nu)$ at $\sqrt{s}=8$ TeV.
 Accordingly, the experimental situation can be applied to  our  $S_{1B,2B}$ schemes. 
  Thus, instead of performing an event simulation, we directly apply the constraints of the ATLAS and CMS measurements to the $S_{1B,2B}$ schemes when $m_t > m_{H^\pm} + m_b$ is satisfied.  We note that although a stricter limit on $m_{H^\pm}$ at $\sqrt{s}=13$ TeV through $H^\pm \to \tau \nu$  was obtained by ATLAS~\cite{Aaboud:2016dig}, the mass range was for $m_{H^\pm} \geq 200$ GeV. Hence, we use the earlier results~\cite{Aad:2014kga,Khachatryan:2015qxa} to constrain the parameters.  
According to the Yukawa couplings in Eq.~(\ref{eq:qyu}), the decay width for $t \to H^+ b$ can be easily formulated as:
\begin{align}
& \Gamma_{t \to H^+ b} = \frac{m_t \sin^2 \theta_\pm}{16 \pi} \sqrt{\lambda(x_{H^\pm},x_b)} \left[ (C_L^2+C_R^2) \sqrt{\tilde \lambda(x_{H^\pm},x_b)} + 4 C_L C_R \frac{m_b}{m_t} \right]\,, \label{eq:tHb} \\
& C_L = \frac{m_t }{v t_\beta }\,, \quad C_R = \frac{t_\beta m_b }{v}\,,  \quad
\tilde \lambda(x,y) = 1+x^2+y^2 -2x +2y -2xy \,, \nonumber 
\end{align}
in which  $V_{tb}\approx 1$ is used, and where $x_{H^\pm} = m_{H^\pm}^2/m_t^2$, and $x_b = m_b^2/m_t^2$. To estimate the BR, the total top-quark decay width was taken as $\Gamma_t =1.41$ GeV~\cite{Agashe:2014kda}. Consequently, we show the  $BR(t \to H^+ b)BR(H^+ \to \tau^+ \nu)$  as a function of $m_{H^\pm}$ in Fig.~\ref{fig:tdecay}, where  the ATLAS (squares) and CMS (triangles) upper limits  are shown in the plots. It should be noted that  the  curves represent the different values of $s_\pm$, and for comparison, we plot the cases with $t_\beta=1$ (left panel) and $t_\beta=30$ (right panel).  From the results, it can be seen that much of the parameter space of $s_\pm$ was excluded by the current LHC data, with 
 the constraint on the case with $t_\beta =30$ being   stronger due to the fact that $BR(H^+ \to \tau^+ \nu)\approx 1$. In the following event simulation,  we thus concentrate on the  $S_{1A,2A}$ schemes. 

To estimate the ratio of  the signal event number ($N_S$) to the background event number $N_B$, we have to determine the SM backgrounds that can mimic the signals of the charged Higgs.  The main background processes are listed  as follows:
\begin{enumerate}
 \item ZZ background: $p p \rightarrow Z Z + n$ jets\,, 
\item WW background: $p p \rightarrow W^\pm W^\mp + n$ jets\,, 
\item WZ background: $p p \rightarrow W^\pm Z + n$ jets \,, 
\item top background: $p p \rightarrow t \bar{t}, t \bar t q(\bar q)$, $t \bar t W^\pm$\,,    
\end{enumerate}
where the number of jets is assumed to be  $n \leq 2$. In order to suppress the background events, we consider the kinematic cuts (KCs), which are applied to all signals and backgrounds,  as:
\begin{align}
\label{eq:cuts_basic}
& p_T (\ell) > 20 \ {\rm GeV}\,, \quad \eta(\ell) < 2.5 \,, \quad 
 p_T(j_{\text{leading}}) > 50 \ {\rm GeV}\,, \nonumber  \\
 &  p_T (j) > 20 \ {\rm GeV}, \quad \eta(j) < 5.0, 
\end{align}
where $p_T$ is the transverse momentum, $\eta = 1/2 \ln (\tan \theta/2)$ is the pseudo-rapidity with $\theta$ being the scattering angle in the laboratory frame, and $j_{\text{leading}}$ denotes the highest $p_T$ jet. Further cuts can be proposed depending on the properties of each process.  In the following, we discuss the simulation analysis for each scheme.

  Scheme  $S_{1A}$: according to the results as shown in Fig.~\ref{fig:BRCH}(b),  it can be known that  $H^+ \to W^+ Z /\bar b bW^+$ are the two main decay channels at small values of $s_\pm$. If we take $m_{H^\pm} =175$ GeV, the  signal processes of interest are
\begin{align}
pp \to H^+ \bar t ( H^+ \bar t b), \quad H^+ \to W^+ Z \, \text{ or } \, \bar b b W^+\,,
\end{align}
where CP-conjugated processes are also implied. 
 To display the signal events for the  $H^\pm \to  b \bar b W^\pm$ and $H^\pm \to W Z$ channels, we set the extra event-selection conditions to be "2 leptons + $n$ jets" and "3 leptons + $m$ jets" with $n(m) \geq 4(3)$, respectively. 
 And,  to increase the significance of  the signal from  the $H^\pm \to W Z$ channel, we further require the  invariant mass of  the 3-lepton to satisfy the selection condition: 
\begin{equation}
120 \, {\rm GeV} \leq M_{\ell^\pm \ell^+ \ell^-} \leq 200 \, {\rm GeV}.
\label{eq:IMcut}
\end{equation}

To clearly show the signal and each background, we present the event numbers after imposing the proposed KCs and extra selection conditions in Table~\ref{tab:Nevents}(\ref{tab:Nevents2}) for the  $H^\pm \to W Z$ ($H^\pm \to \bar b b W^\pm$) signal and background, where the employed  luminosity is 100 fb$^{-1}$. Note that  the last column is the significance, which is defined as $S=N_S/\sqrt{N_B}$.  Since the BR of the $W Z$ mode can be higher than  that of the $\tau \nu$ mode at small $s_\pm$, we take $s_\pm=0.2$ for the  $H^\pm \to W Z$ channel. However, in order to obtain a larger $H^\pm$ production cross section, we take $s_\pm =0.4$ for the  $H^\pm \to  \bar b b W^\pm$ channel.  From Table~\ref{tab:Nevents}, it can be seen that with the proposed cuts, the background events from the top-quark and $WZ$ backgrounds are still much larger than the  signal events. In principle, we can further reduce the top-quark background by imposing the KC on the second highest $p_T$ of the charged lepton; however, doing so does not significantly change the  $WZ$ background. The reason for this  is that the kinematic distributions from the  $H^\pm \to W Z$ signal and from the $WZ$ background are very similar.  As such, the proposed KC will reduce both events. A similar situation also occurs, as shown  in Table~\ref{tab:Nevents2}, for the signal from the $H^\pm \to \bar b b W^\pm$ channel. The difference is that we can find a KC (e.g., invariant mass of $\ell^+ \ell^-$) to reduce the $WZ$ and $ZZ$ backgrounds; nevertheless, due to the similarity in the kinematic distributions between the $H^\pm \to  \bar b b W^\pm$ signal and  top-quark background,  finding  an efficient cut that diminishes the top-quark background without reducing the signal is challenging.

From the results shown in both tables, it can be clearly seen that the significance from $H^\pm \to W Z$ is much smaller than that from $H^\pm \to b \bar b W^\pm$.  The results can be easily understood as follows: the product of $\sigma(pp\to H^+ t, H^+ b \bar t) BR(H^+ \to W^+ Z)$ at  $s_\pm =0.2$ is close to that at $s_\pm =0.4$; however,  $\sigma(pp\to H^+ t, H^+ b \bar t)BR(H^+ \to b\bar b W^+)$ at  $s_\pm =0.4$ is  30-fold larger than with $H^+ \to W^+Z$. By the definition of significance, the  $S$ of $H^\pm \to  W Z$ natively should be one order of magnitude smaller than that of $H^\pm \to  \bar b b W^\pm$.  In sum, by combining all analyses, we conclude that it is difficult to search for a light -charged Higgs via the $H^\pm \to W Z$ channel. 


\begin{table}[hptb]
\begin{ruledtabular}
\begin{tabular}{lccccccc}
cuts & signal(3$\ell$) & $WW$+$n$\,j & $ZZ$+$n$\,j & $WZ$+$n$\,j  & top & top+$W$ & S \\ \hline \hline
KCs & 26. & 27. & 8.9$\times 10^2$ & 8.8$\times 10^3$ & $5.8 \times 10^3$  & $1.1 \times 10^2$  & 0.21 \\ \hline
$M_{\ell^\pm \ell^+ \ell^-}$ cut & 19. & $9.0$ & 40. & $5.1 \times 10^3$ & $3.2 \times 10^3$ & 50. &  0.15
\end{tabular}
\caption{Event number for the  $H^\pm \to W Z$  signal and background with the proposed kinematic cuts in the scheme $S_{1A}$, where we have used the luminosity of 100 fb$^{-1}$, $m_{H^{\pm}} =175$ GeV, and $s_\pm =0.2$.  \label{tab:Nevents}}
\end{ruledtabular}
\end{table}

\begin{table}[hptb]
\begin{ruledtabular}
\begin{tabular}{lccccccc}
cuts & signal(2$\ell$) & $WW$+$n$\,j & $ZZ$+$n$\,j & $WZ$+$n$\,j  & top & top+$W$ & S \\ \hline \hline
KCs &  $2.3 \times 10^3$ & $2.4\times 10^4$ & $1.9 \times 10^4$ & $5.5 \times 10^4$ & $8.1 \times 10^5$  & $1.6 \times 10^3$ & 2.4 
\end{tabular}
\caption{ This legend is the same as that in Table~\ref{tab:Nevents}, except here for the  $H^\pm \to b \bar b W^\pm$ signal and $s_\pm=0.4$.  \label{tab:Nevents2}}
\end{ruledtabular}
\end{table}

Scheme $S_{2A}$ :  in this scheme,   it can be found from the plot shown in Fig.~\ref{fig:BRCH}(a) that the charged Higgs mainly decays to $\tau \nu$ and $\bar b b W^\pm$, where the decay priority depends on $m_{H^\pm}$. For the case where $\tau \nu$ dominates, we can  use the ATLAS and CMS measurements directly to constrain the parameters,  as discussed earlier, for which  the results are shown in Fig.~\ref{fig:tdecay}. We thus focus the simulation on the $\bar b b W^\pm$ channel by selecting some benchmark points (BPs) for the  parameters instead of scanning all  parameter spaces.
 Since the analysis is similar to that for  scheme $S_{1A}$, we directly present the signal event numbers and significances with the selected BPs of $(m_{H^\pm}, s_\pm)$ in Table.~\ref{tab:Nevents3}, and assumed the same luminosity, KCs, and backgrounds as those in Table~\ref{tab:Nevents2}.  In addition, the selected BPs  satisfy the ATLAS~\cite{Aad:2014kga} and CMS~\cite{Khachatryan:2015qxa}  upper limits  shown in Fig.~\ref{fig:tdecay}. From the table, it can be seen that a heavier $H^\pm$ and larger $s_\pm$ exhibit  a greater significance. This is because  when  the $H^\pm$ boson becomes heavier, in addition to the $BR(H^\pm \to \bar b b W^\pm)$ approaching unity,  the larger allowed $s_\pm$ values lead to larger single $H^\pm$ production cross-sections. 
 
\begin{table}[hptb]
\begin{ruledtabular}
\begin{tabular}{lccc}
$(m_{H^\pm}[{\rm GeV}],\sin \theta_\pm)$ & (150, \, 0.2) & (120, \, 0.1) & (100, \, 0.1)  \\ \hline 
\# of events &  $5.9 \times 10^3$ & $1.2 \times 10^3$ &  $6.4 \times 10^2$ \\ \hline
$S$ & $6.2$ & $1.3$ & $0.67$
\end{tabular}
\caption{ Number of signal events and the associated significance with some selected benchmark points for the $H^\pm \to \bar b b W^\pm$ signal, where the luminosity, kinematic cuts and backgrounds are the same as those in Table~\ref{tab:Nevents2}. The selected benchmark  points satisfy the ATLAS~\cite{Aad:2014kga} and CMS~\cite{Khachatryan:2015qxa}  upper limits on $BR(t\to H^+ b)BR(H^+ \to \tau^+ \nu)$.  \label{tab:Nevents3}}
\end{ruledtabular}
\end{table}

In summary, we assessed the discovery potential of the light-charged Higgs in the two-Higgs-doublet and one-Higgs-triplet model at the center of a mass energy of $\sqrt{s}=13$ TeV. If $H^\pm \to \tau \nu$ is the dominant decay channel, then the current ATLAS~\cite{Aad:2014kga} and CMS~\cite{Khachatryan:2015qxa} upper limits on the product of $BR(t\to H^+ b)BR(H^+ \to \tau^+\nu)$ can be directly used, and the correlation between $m_{H^\pm}$ and $\sin\theta_\pm$ is severely constrained. Although the $H^\pm \to W Z$ channel is allowed,  and its branching ratio is not suppressed in the model, the significance of the   $H^\pm \to W Z$  signal is much smaller than that of  $H^\pm \to  \bar b b W^\pm$. Accordingly, we conclude that  the optimal better production processes to search for a light-charged Higgs are $pp\to H^+ \bar t, H^+ b \bar t$,  for which the charged Higgs decay channel is $H^\pm \to \bar b b W^\pm$.


\section*{Acknowledgments}
This work was partially supported by the Ministry of Science and Technology of Taiwan
R.O.C.,  under grant MOST-103-2112-M-006-004-MY3 (CHC). 


\end{document}